\documentclass[runningheads]{llncs}

\usepackage{graphicx}
\usepackage{url}
\usepackage{todonotes}
\usepackage{multirow}
\usepackage{amsmath}

\PassOptionsToPackage{hyphens}{url}
\usepackage{hyperref}
\expandafter\def\expandafter\UrlBreaks\expandafter{\UrlBreaks
  \do\a\do\b\do\c\do\d\do\e\do\f\do\g\do\h\do\i\do\j%
  \do\k\do\l\do\m\do\n\do\o\do\p\do\q\do\r\do\s\do\t%
  \do\u\do\v\do\w\do\x\do\y\do\z\do\A\do\B\do\C\do\D%
  \do\E\do\F\do\G\do\H\do\I\do\J\do\K\do\L\do\M\do\N%
  \do\O\do\P\do\Q\do\R\do\S\do\T\do\U\do\V\do\W\do\X%
  \do\Y\do\Z\do\*\do\-\do\~\do\'\do\"\do\-}%

\newcommand{\lang}[1]{\texttt{#1}}
\newcommand{\en}{\lang{EN}}
\newcommand{\ru}{\lang{RU}}
\newcommand{\de}{\lang{DE}}

\lccode`\+`\+

\newcommand\dataset[0]{EventKG+Click}

\begin{document}





\title{EventKG+Click: A Dataset of Language-specific Event-centric User Interaction Traces}


\author{Sara Abdollahi \and Simon Gottschalk \and Elena Demidova}

\titlerunning{\dataset{}}
\authorrunning{S. Abdollahi et al.}

\institute{L3S Research Center, Leibniz Universität Hannover, Germany \\
\email{\{abdollahi,gottschalk,demidova\}@L3S.de}}

\maketitle

\begin{abstract}
An increasing need to analyse event-centric cross-lingual information calls for innovative user interaction models that assist users in crossing the language barrier. However, datasets that reflect user interaction traces in cross-lingual settings required to train and evaluate the user interaction models are mostly missing. In this paper, we present the \dataset{} dataset that aims to facilitate the creation and evaluation of such interaction models. \dataset{} builds upon the event-centric EventKG knowledge graph and language-specific information on user interactions with events, entities, and their relations derived from the Wikipedia clickstream.  
\end{abstract}

\section{Introduction} 
\label{sec:introduction}

With a rapidly growing number of events with significant international impact, cross-lingual analytics gains increased importance for researchers and professionals in many disciplines, including digital humanities, media studies, and journalism. The most prominent recent examples of such events include the COVID-19 outbreak, the migration crisis in Europe, and Brexit. From the information science perspective, research on event-centric information spread across languages and communities, as well as cross-cultural and cross-lingual differences in reporting, are of particular interest. However, very often, the language barrier hinders such research.
{\let\thefootnote\relax\footnote{{Copyright © 2020 for this paper by its authors. Use permitted under Creative Commons License Attribution 4.0 International (CC BY 4.0).}}}

The development of novel methods for user interaction with event-centric cross-lingual information can help to overcome the language barrier in this context. Such methods can facilitate researchers with limited knowledge of target languages to narrow down the search space and to obtain an overview of the cross-lingual differences effectively and efficiently. However, currently, user interaction in multilingual settings is not sufficiently studied. The benchmarks and datasets suitable for the evaluation of new methods for user interaction with cross-lingual information are mostly missing.

With the recent development of knowledge graphs that provide cross-lingual information, such as Wikidata, DBpedia, and the event-centric EventKG knowledge graph \cite{gottschalk2019eventkg}, the availability of semantic event-centric cross-lingual information has significantly increased. These knowledge graphs contain semantic information regarding events and their relations while providing labels in different languages along with the properties extracted from language-specific sources. 
For example, EventKG, in its version 2.1 released in February 2020, includes information on more than $1,200,000$ events in nine languages. 
We believe that knowledge graphs containing event-centric cross-lingual data can build a backbone for the development of user interaction methods that can assist users in crossing the language barrier.

In this paper, we present a novel cross-lingual dataset that reflects the lan\-guage-specific relevance of events and their relations. 
This dataset aims to provide a reference source to train and evaluate novel models for event-centric cross-lingual user interaction, 
with a particular focus on the models supported by knowledge graphs. 
Our dataset \dataset{} is based on two data sources: 
1) the Wikipedia clickstream\footnote{\url{https://meta.wikimedia.org/wiki/Research:Wikipedia_clickstream}} that reflects real-world user interactions with events and their relations within language-specific Wikipedia editions; and 
2) the EventKG knowledge graph that contains 
semantic information regarding events and their relations that partially originates from Wikipedia.
\dataset{} is available online\footnote{\url{https://github.com/saraabdollahi/EventKG-Click}} to enable further analyses and applications.

Without loss of generality, we adopt a language-specific event ranking as an envisioned user interaction paradigm to illustrate our discussion.
For example, Table \ref{tab:entity_relevance} reveals the different language-specific focus when ranking events.
In each of the three languages contained in \dataset{}, the list of most language-specific related events is clearly representing language-specific views (e.g., ``2016 Berlin truck attack'' for German) that can be used for further exploration of events from language-specific viewpoints. In the case of English, we see that the Southeast Asian Games are of high language-specific relevance, which can be explained by the large percentage of Asian users of the English Wikipedia\footnote{\url{https://stats.wikimedia.org/wikimedia/squids/SquidReportPageViewsPerCountryBreakdown.htm}}.

\begin{table}
\center
\caption{Events with highest language-specific relevance per language in \dataset{}.}
\label{tab:entity_relevance}
\begin{tabular}{|c|l|l|l|}
\hline
\multicolumn{1}{|l|}{\textbf{Rank}} & \multicolumn{1}{c|}{\textbf{English}} & \multicolumn{1}{c|}{\textbf{German}} & \multicolumn{1}{c|}{\textbf{Russian}} \\ \hline
\textbf{1} & \begin{tabular}[c]{@{}l@{}}Southeast\\ Asian Games\end{tabular} &  \begin{tabular}[c]{@{}l@{}}2016 Berlin\\ truck attack\end{tabular} & \begin{tabular}[c]{@{}l@{}}2009 Russian\\ Premier League\end{tabular} \\ \hline
\textbf{2} & \begin{tabular}[c]{@{}l@{}}2017 Southeast\\ Asian Games\end{tabular} &  \begin{tabular}[c]{@{}l@{}}German student\\ movement \end{tabular} & \begin{tabular}[c]{@{}l@{}}1993 Russian\\ Top League\end{tabular} \\ \hline
\textbf{3} & \begin{tabular}[c]{@{}l@{}}2014 United States\\ Senate elections\end{tabular} &  \begin{tabular}[c]{@{}l@{}}2006 Austrian\\ legislative election\end{tabular} & \begin{tabular}[c]{@{}l@{}} 2012–13 Russian\\ Premier League\end{tabular} \\ \hline

\end{tabular}
\end{table}

In \dataset{}, we enrich the information obtained from the Wikipedia clickstream with event and entity references from EventKG. 
Furthermore, we create a cross-lingual view on the clickstream by combining information obtained from three Wikipedia language editions, namely English, German, and Russian. 
Moreover, we compute scores that reflect the language-specific relevance of events and their relations, as indicated by the user interactions in the clickstream. 
Finally, to support further development of the event-centric user interaction methods in the cross-lingual settings, 
we analyse the correlations of the proposed scoring function and selected influence factors.  
%

We structure the rest of the paper as follows: First, we review related work regarding cross-lingual analytics, knowledge graphs, and the Wikipedia clickstream in Section \ref{sec:related}. Then, we introduce our \dataset{} dataset in Section \ref{sec:datasets}. In Section \ref{sec:lang-spec-rel}, we propose scores to represent the language-specific relevance of events and their relations. Given the \dataset{} dataset and these scores, we analyse how selected factors influence the language-specific relevance in Section \ref{sec:factors}. Finally, we provide a conclusion in Section \ref{sec:conclusion}.

\section{Related Work}
\label{sec:related}

In this section, we briefly summarise related work in the areas of cross-lingual analytics, along with the aspects related to knowledge graphs and the Wikipedia clickstream.

\textbf{Cross-lingual analytics and interaction.} With the rise of the Web, there came an uprise of user-generated content accessible over the whole world, leading to knowledge diversity across languages \cite{hecht2010tower}. The identification and analysis of such knowledge diversity is an important method to understand language communities better. For example, Oeberst et al. identified different types of "collective biases" such as biased representations of intergroup conflicts that appear under collaborative circumstances \cite{oeberst2016individual}. Miz et al. identified how Wikipedia reflects cultural particularities \cite{miz2020trending}. Mocanu et al. have identified linguistic trends in Twitter usage in more than 100 countries \cite{mocanu2013twitter}.

In the context of cross-lingual analytics, events play a particularly important role: When an event breaks out, this event is usually reported by a large number of sources, whose coverage highly varies across language communities \cite{el2018measuring}. This phenomenon becomes visible when using EventRegistry, a tool that allows cross-lingual exploration of news articles which are assigned to event clusters \cite{rupnik2016news}. Event-centric cross-lingual analytics are also viable across different Wikipedia language editions as illustrated by two case studies about the Brexit and the US withdrawal from the Paris Agreement, where researchers identified language-specific viewpoints \cite{gottschalk2018towards}.

With \dataset{}, our goal is to promote further cross-lingual analytics and interaction, facilitated by a combination of semantic information given in knowledge graphs and user interaction traces obtained from a clickstream.

\textbf{Knowledge graphs.} An essential resource to facilitate interaction with cross-lingual information are knowledge graphs, in particular those containing language-specific labels and relations. Kaffee et al. \cite{kaffee2019ranking} developed metrics that measure the multilingualism of knowledge graphs to identify those suitable for usage in multilingual applications and to gain cross-lingual insights. For example,  Marie et al. \cite{marie2014exploratory} discovered a "cultural prism'' between the different DBpedia language editions when querying for entities related to facets of interest.

The importance of multilingualism in knowledge graphs becomes even more evident in the case of event-based applications. EventKG \cite{gottschalk2019eventkg} is a knowledge graph that is tailored not only to the interaction with event-centric information but also contains information coming from several languages. An example application that makes use of this cross-lingual event knowledge is EventKG+TL \cite{gottschalk2018eventkgtl} that relies on Wikipedia link counts present in EventKG to model the importance of events related to a given concept.

In our analysis, we observed that the closeness of event locations extracted from EventKG is an essential indicator to explain language-specific relevance. Thus, we confirm the importance of event-centric and multilingual knowledge graphs in the context of cross-lingual analytics.

\textbf{Wikipedia clickstream.}  
The Wikipedia clickstream has been used as a ground-truth to evaluate entity recommendation and relatedness in several examples, as it reveals the navigationâl behaviour of users and their preferences while exploring  Wikipedia pages. 
Existing work, however, has not considered language-specific differences and mainly focused on the English Wikipedia clickstream: For example, Tran et al. used the English Wikipedia clickstream as ground truth for constructing entity-context queries \cite{tran2017beyond} and Bhatia et al. constructed their query dataset based on the English Wikipedia clickstream \cite{bhatia2018know}. Nguan et al. evaluated their relatedness ranking method by using the raw number of navigations in Wikipedia clickstream \cite{nguyen2018trio}.
With the usage of the Wikipedia clickstream in different languages, \dataset{} adds a new perspective onto EventKG, as it reflects real user behaviour across language communities, which goes beyond the consideration of knowledge graph relations and Wikipedia link counts.

\section{\dataset{} Dataset}
\label{sec:datasets}

The Wikipedia clickstream holds the interaction of real users with the articles representing events and entities in the specific Wikipedia language editions and their relations. In particular, the clickstream contains the counts of the (source, target) pairs extracted from Wikipedia's request logs. The clickstream contains all the requests to a Wikipedia page, including links from and to external web pages. As \dataset{} and our analysis are based on Wikipedia click behaviour, we only consider those (source, target) click pairs in the clickstream where both the source and target are Wikipedia articles connected by a hyperlink.

In this work, we adopt the Wikipedia clickstream that covers the period from December 1, 2019, to December 31, 2019,
and contains nearly $19,521,580$ click pairs for the English, $2,902,878$ click pairs for the German, and $2,752,340$ click pairs for the Russian Wikipedia.

EventKG is an event-centric knowledge graph that contains more than $1.2$ million events and more than $4$ million temporal relations in nine languages in its release from February 2020.
Knowledge graphs such as EventKG, DBpedia, and Wikidata include information extracted from the multilingual Wikipedia as the basis. This way, data regarding user interaction with Wikipedia articles and links, 
available from the Wikipedia clickstream dataset, can be directly mapped to the events, entities and their relations in these knowledge graphs. 

When creating the proposed \dataset{} dataset, we assume that: 
1) the events of global importance are reflected in Wikipedia clickstreams of several languages, and
2) a clickstream in a specific language reflects the importance of events and their relations as perceived by the users of the specific Wikipedia language edition.
Based on these assumptions, we employ the intersection of language-specific clickstreams to build a dataset for training and evaluation of cross-lingual user interaction. 
In particular, we map the events and entities included in the Wikipedia clickstream to EventKG and extract relations for these events from all language-specific clickstreams. 
Furthermore, we compute scores that represent the language-specific relevance of events and their relations. These scores are presented in Section \ref{sec:lang-spec-rel}. To enable further cross-lingual analysis, we enrich \dataset{} with several influence factors extracted from EventKG and Wikipedia, which are presented in Section \ref{sec:factors}.

In \dataset{}, we only consider entities that are clicked at least $10$ times per language, so that we capture those entities that are of global importance and do not consider entities solely present in single Wikipedia language versions. We also only consider pairs which exist in the clickstreams of all considered languages and in which the target page is an event. 

The resulting \dataset{} dataset is available online\footnote{\url{https://github.com/saraabdollahi/EventKG-Click}} and contains relevance scores for 
more than $4$ thousand events, and nearly $10$ thousand event-centric click-through pairs.

\section{Scores to Assess Language-Specific Relevance}
\label{sec:lang-spec-rel}

To allow cross-lingual analytics with \dataset{}, we need to capture the language-specific relevance of events and their relations. Based on the Wikipedia clickstream, we propose two scores that rule out language-independent relevance.

To describe our scores, we first define the concepts used for the computation:

\begin{itemize}
\item $L$ is the set of languages under consideration. The current release of \dataset{} comes in English, German, and Russian: $L=\{EN,DE,RU\}$.
\item $E$ is the set of entities contained in \dataset{}, that are all represented by specific Wikipedia pages and EventKG resources. Formally, named events considered in this work are a specific type of entity and thus included in $E$.
\item $clicks(e_s,e_t,l)$ represents the number of clicks from the source entity $e_s \in E$ to the target event $e_t \in E$ in the clickstream of the given language $l \in L$.
\end{itemize}

We distinguish between two scores defined in the following: language-specific event relevance and language-specific relation relevance. 

\subsection{Language-specific Event Relevance}

Wikipedia language versions differ a lot concerning the number of their active users, edits, and articles. For example, the English Wikipedia has $7.2$ times as many active users as the German Wikipedia\footnote{\url{https://en.wikipedia.org/wiki/List_of_Wikipedias}}. The clickstream also reflects this imbalance: There are $7$ times more clicks in the English clickstream than in the German one. To observe language-specific behaviour, we first need to level the effects that originate from the popularity of the specific Wikipedia language versions.
To do so, we normalise the number of clicks with respect to the total number of clicks in the respective language, which leads to normalised scores in the range $[0,1]$. In order to create balanced click counts, we then multiply the normalised score by the total number of clicks in the clickstreams, as follows:

\begin{equation*}
balanced\_clicks(e_s,e_t,l) = clicks(e_s,e_t,l) \cdot \frac{\sum_{l' \in L}\sum_{e_s' \in E}\sum_{e_t' \in E} clicks(e_s',e_t',l')}{\sum_{e_s' \in E}\sum_{e_t' \in E} clicks(e_s',e_t',l)}
\end{equation*}

The popularity of an event can be inferred by the number of user interactions with its Wikipedia page. That way, we can identify the most popular events in a given language $l \in L$ by summing up all clicks from and to an event $e \in E$:

\begin{equation*}
balanced\_clicks(e,l) = \sum_{e_t \in E}balanced\_clicks(e,e_t,l)+\sum_{e_s \in E}balanced\_clicks(e_s,e,l)
\end{equation*}

As we focus on the language-specific relevance in \dataset{}, we need to rule out the events that are highly ranked across all languages under consideration. Therefore, we normalise the language-specific click count by the overall number of clicks in all languages:

\begin{equation*}
event\_relevance(e,l) =  \frac{balanced\_clicks(e,l)}{\sum_{l' \in L} balanced\_clicks(e,l')} \in [0,1]
\end{equation*}

With this relevance score, events that are clicked often in a given language $l \in L$, but rarely clicked in the other languages are assigned a relevance score close to $1$.

\subsection{Language-specific Relation Relevance}

To identify events relevant to a given source entity, we define the language-specific relation relevance score. This score assigns a relevance score to the relation between a source entity $e_s$ and a target event $e_t$ in a given language. Similarly to the language-specific event relevance, the language-specific relation relevance is computed as the fraction of clicks in the given language compared to all languages:

\begin{equation*}
relation\_relevance(e_s,e_t,l) = \frac{balanced\_clicks(e_s,e_t,l)}{\sum_{l' \in L} balanced\_clicks(e_s,e_t,l')} \in [0,1]
\end{equation*}

Note that this score rules out the effects resulting from the relevance of the source entity: Events that are highly related to an entity $e$ can obtain relevance scores close to $1$ independent of $e$'s click count.

\subsection{Examples of Scores}

In Table \ref{tab:entity_relevance} in Section \ref{sec:introduction}, we have given an example of the language-specific event relevance, i.e., that table provides the top-ranked events per language, according to our language-specific event relevance score. As discussed before, we can clearly observe events which are intuitively important for the respective language community.

\begin{table}
\center
\caption{The three events most relevant to the 2012 Summer Olympics for English, German and Russian.}
\label{tab:relation_relevance}
\begin{tabular}{|c|l|l|l|}
\hline
\textbf{Rank} & \multicolumn{1}{c|}{\textbf{English}} & \multicolumn{1}{c|}{\textbf{German}} & \multicolumn{1}{c|}{\textbf{Russian}} \\ \hline
\textbf{1} & \begin{tabular}[c]{@{}l@{}}2012 Summer Olympics\\ opening ceremony\end{tabular} & \begin{tabular}[c]{@{}l@{}}Olympic Games\end{tabular} & \begin{tabular}[c]{@{}l@{}}2020 Summer Olympics\\ opening ceremony\end{tabular} \\ \hline
\textbf{2} & \begin{tabular}[c]{@{}l@{}}Swimming at the\\ 2012 Summer Olympics\end{tabular} & \begin{tabular}[c]{@{}l@{}}Modern pentathlon at the \\ 2012 Summer Olympics\end{tabular} & \begin{tabular}[c]{@{}l@{}}Olympic Games\end{tabular} \\ \hline
\textbf{3} & \begin{tabular}[c]{@{}l@{}}Badminton at the \\ 2012 Summer Olympics\end{tabular} & \begin{tabular}[c]{@{}l@{}}Equestrian at the\\ 2012 Summer Olympics\end{tabular} & \begin{tabular}[c]{@{}l@{}}Weightlifting at the\\ 2012 Summer Olympics\end{tabular} \\ \hline
\end{tabular} 
\end{table}

Table \ref{tab:relation_relevance} presents the language-specific relation relevance by showing the concrete example of events relevant to the Summer Olympics in 2012. According to our score, the opening ceremony that happened in London is the most relevant event for the 2012 Summer Olympics from the English perspective. Apart from that, we can observe that sports particularly popular in a specific language community are ranked higher (e.g., swimming for English, equestrian sports for German, and weightlifting for Russian).

Our examples illustrate that user click behaviour is not only based on globally relevant entities but takes the language-specific relevance into account. Both relevance scores can be used in language-specific contexts, e.g. for event retrieval or recommendation.

\section{Influence Factors for Language-Specific Relevance}
\label{sec:factors}

Given the \dataset{} dataset with the relevance scores defined in the previous section, we now discuss several \textit{influence factors} that can potentially impact the language-specific relevance of events and analyse their correlations with the proposed relevance scores. As 
influence factors we consider language community relevance, event location closeness and event recency, as defined in the following. In future work, we plan to investigate the role of further influence factors, as for example the event type that has been shown to influence the click-behaviour \cite{dimitrov2018query}.

\subsection{Language Community Relevance}

The language community relevance factor reflects the importance of an event for the community that speaks this language. We assume that events relevant for the language community should be mentioned and referred to more often in a language-specific corpus.

Based on this assumption, we measure the language community relevance by counting the links to the event article and mentions of the event within the specific Wikipedia language edition\footnote{We derive these counts from EventKG that contains link and mention counts \cite{gottschalk2019eventkg}.}. 
Dependent on the context (i.e., event or relation relevance), we make use of two influence factors:
\begin{itemize}
    \item Links pointing to the event: The number of links in the whole Wikipedia language edition that link to the event article.
    \item Co-mentions of a relation: The number of sentences in the whole Wikipedia language edition that jointly mentions the (source, target) pair participating in the relation.
\end{itemize}

\subsection{Event Location Closeness}

The event location closeness factor expresses the intuition that users are likely to be interested in the exploration of local events, i.e., events located in spatial proximity of the user. 
To reflect this intuition, we introduce a binary influence factor that indicates whether an event happened in a location where the respective language $l \in L$ is an official language. For example, the Battle of Stalingrad may be particularly important from the Russian perspective in the context of the Second World War. 
To compute this factor, we first identify event location(s) using the \texttt{sem:hasPlace}\footnote{\url{http://semanticweb.cs.vu.nl/2009/11/sem/hasPlace}} property of EventKG and then derive the official languages of the location's country.

\subsection{Event Recency}

Wikipedia is heavily influenced by recent events: Users tend to edit and read articles about events that are happening right now \cite{kaltenbrunner2012there}. To observe the impact of recency on the language-specific user click behaviour, we introduce a recency score, which is computed as the number of days between the event start date and the start date of the clickstream dataset (the dates of the specific entries in the dataset are not available). To identify the event start dates, we use \texttt{sem:hasBeginTimeStamp} values in EventKG.

\subsection{Correlations with Influence Factors}

Given \dataset{} and the influence factors, we now investigate the correlations between such influence factors and the language-specific relevance scores. To this end, we compute the Pearson correlation coefficients in several configurations.

First, we compute the correlations of influence factors with language-specific event relevance scores of the events covered in the Wikipedia clickstream of all considered languages (i.e., $event\_relevance$, as defined in Section \ref{sec:lang-spec-rel}). As influence factors we select the event location closeness (\textit{Location}), the number of links pointing to the respective event (\textit{Links}), and the event recency (\textit{Recency}). Results are shown in Table \ref{tab:entity_correlations}.

\begin{table}
\center
\caption{Correlations of influence factors with event relevance scores in \dataset{}.}
\label{tab:entity_correlations}
\begin{tabular}{cc|r|r|r|r|r|r|r|}
\cline{3-9}
\multicolumn{1}{l}{} & \multicolumn{1}{l|}{} & \multicolumn{7}{c|}{\textbf{Influence Factors}} \\ \cline{3-9} 
\multicolumn{1}{l}{} & \multicolumn{1}{l|}{} & \multicolumn{3}{c|}{\textbf{Location}} & \multicolumn{3}{c|}{\textbf{Links}} & \multirow{2}{*}{\textbf{\begin{tabular}[c]{@{}c@{}}Recency\end{tabular}}} \\ \cline{3-8} 
\multicolumn{1}{l}{} & \multicolumn{1}{l|}{} & \multicolumn{1}{c|}{\textbf{\en}} & \multicolumn{1}{c|}{\textbf{\de}} & \multicolumn{1}{c|}{\textbf{\ru}} & \multicolumn{1}{c|}{\textbf{\en}} & \multicolumn{1}{c|}{\textbf{\de}} & \multicolumn{1}{c|}{\textbf{\ru}} & \\ \hline \cline{2-9}
\multicolumn{1}{|c|}{\multirow{3}{*}{\textbf{\begin{tabular}[c]{@{}c@{}}Language-\\ specific\\ relevance\end{tabular}}}} & \textbf{\en} & 0.4 & -0.13 & -0.25 & 0  & -0.02 & 0.01 & -0.19  \\ \cline{2-9} 
\multicolumn{1}{|c|}{} & \textbf{\de} & -0.18 & 0.20 & 0.02 & -0.01 & 0 & 0.01 & 0.12 \\ \cline{2-9} 
\multicolumn{1}{|c|}{} & \textbf{\ru} & -0.19 & -0.08 & 0.26 & -0.01  & -0.02 & 0.03 & 0.07  \\ \hline
\end{tabular}
\end{table}

The \textit{Location} influence factor for events indicates the largest positive correlation, which confirms the existence of different language viewpoints. This effect can be most notably observed in the case of English, which has a correlation of $0.4$ between the event relevance score and the \textit{Location} closeness influence factor. 
The other two influence factors, namely \textit{Links} and \textit{Recency}, do not show any notable correlation. 
We assume that this is because the users are interested in both, recent and historical events, whereas recent events might not be well interlinked in Wikipedia yet.

Until now, we have considered the language-specific event relevance scores, i.e., scores assigned to each event in isolation. Now, we investigate the user click behaviour from the perspective of the event relations (i.e., $relation\_relevance$, as defined in Section \ref{sec:lang-spec-rel}). 
In particular, we focus on the properties of the target event, as the language-specific relation relevance score is independent of the source entity's relevance. 

The following influence factors are used in this correlation analysis: 
\begin{itemize}
\item [i] \textit{Location}: The location closeness of the target event. 
\item [ii] \textit{Links}: The number of links to the target event in Wikipedia.
\item [iii] \textit{Recency}: The recency of the target event.
\item [iv] \textit{Co-Mentions}: The number of co-mentions of the relation source and target in Wikipedia.
\end{itemize}

\begin{table}
\center
\caption{The Pearson correlation coefficient of the relation relevance score with selected influence factors in \dataset{}.}
\label{tab:relation_correlations}
\begin{tabular}{cc|r|r|r|r|r|r|r|r|r|r|}
\cline{3-12}
\multicolumn{1}{l}{} & \multicolumn{1}{l|}{} & \multicolumn{10}{c|}{\textbf{Influence Factors}} \\ \cline{3-12} 
\multicolumn{1}{l}{} & \multicolumn{1}{l|}{} & \multicolumn{3}{c|}{\textbf{Location}} & \multicolumn{3}{c|}{\textbf{Links}} & \multirow{2}{*}{\textbf{\begin{tabular}[c]{@{}c@{}}Recency\end{tabular}}} & \multicolumn{3}{l|}{\textbf{Co-mentions}} \\ \cline{3-8} \cline{10-12} 
\multicolumn{1}{l}{} & \multicolumn{1}{l|}{} & \multicolumn{1}{c|}{\textbf{\en}} & \multicolumn{1}{c|}{\textbf{\de}} & \multicolumn{1}{c|}{\textbf{\ru}} & \multicolumn{1}{c|}{\textbf{\en}} & \multicolumn{1}{c|}{\textbf{\de}} & \multicolumn{1}{c|}{\textbf{\ru}} & & \multicolumn{1}{c}{\textbf{\en}} & \multicolumn{1}{c|}{\textbf{\de}} & \multicolumn{1}{c|}{\textbf{\ru}} \\ \hline
\multicolumn{1}{|c|}{\multirow{3}{*}{\textbf{\begin{tabular}[c]{@{}c@{}}Language-\\ specific\\ relevance\end{tabular}}}} & \textbf{\en}  & 0.41 & -0.09 & -0.27 & 0 & 0 & -0.01 & -0.16  & 0.11 & -0.06 & -0.04 \\ \cline{2-12} 
\multicolumn{1}{|c|}{} & \textbf{\de}  & -0.13 & 0.10 & 0.01 & 0 & 0  & -0.01 & 0.1 & -0.01 & 0 & 0.01 \\ \cline{2-12} 
\multicolumn{1}{|c|}{} & \textbf{\ru} &  -0.15 & -0.05 & 0.18 & 0 & 0 & 0.02 & 0.06 & -0.07 & -0.07 & 0.13 \\ \hline
\end{tabular}
\end{table}


The correlation results are shown in Table \ref{tab:relation_correlations}. The correlation coefficient for the language-specific relation relevance confirms our observations concerning the language-specific event relevance. 
The closeness of the target event location has the largest influence on language-specific relevance. 
The links, recency and co-mentions do not correlate with the relevance scores in any of the three languages. 
That means, if the user reads a particular Wikipedia article, there is a higher chance that the next click leads to a spatially close event than to an event that is mentioned many times together with the source entity.

\section{Conclusion and Outlook}
\label{sec:conclusion}

In this paper, we presented the \dataset{} dataset and suggested scores for capturing language-specific relevance scores for events and their relations. \dataset{} builds upon the EventKG knowledge graph and language-specific traces of user interaction with events derived from the Wikipedia clickstream. 
The resulting \dataset{} dataset contains click counts and relevance scores for more than $4$ thousand events and more than $10$ thousand (source, target) pairs in English, German, and Russian. 
Furthermore, we analysed several influence factors of language-specific relevance.  
We believe that the \dataset{} dataset is a valuable resource to evaluate event relevance in language-specific contexts.
In future work, we plan to develop novel user interaction models supporting cross-lingual event-centric analytics, where we will adopt the \dataset{} dataset for training and evaluation.

\subsubsection*{Acknowledgements} 
This work was partially funded by H2020-MSCA-ITN-2018-812997 under ``Cleopatra''.

\bibliographystyle{splncs04}
\bibliography{bibliography.bib}

\begin{thebibliography}{10}
\providecommand{\url}[1]{\texttt{#1}}
\providecommand{\urlprefix}{URL }
\providecommand{\doi}[1]{https://doi.org/#1}

\bibitem{bhatia2018know}
Bhatia, S., Vishwakarma, H.: {Know thy Neighbors, and More! Studying the Role
  of Context in Entity Recommendation}. In: Proceedings of the 29th on
  Hypertext and Social Media. pp. 87--95 (2018)

\bibitem{dimitrov2018query}
Dimitrov, D., Lemmerich, F., Fl{\"o}ck, F., Strohmaier, M.: {Query for
  Architecture, Click through Military: Comparing the Roles of Search and
  Navigation on Wikipedia}. In: Proceedings of the 10th ACM Conference on Web
  Science. pp. 371--380 (2018)

\bibitem{el2018measuring}
El~Ali, A., Stratmann, T.C., Park, S., Sch{\"o}ning, J., Heuten, W., Boll,
  S.C.: {Measuring, Understanding, and Classifying News Media Sympathy on
  Twitter after Crisis Events}. In: Proceedings of the 2018 CHI Conference on
  Human Factors in Computing Systems. pp. 1--13 (2018)

\bibitem{gottschalk2018towards}
Gottschalk, S., Bernacchi, V., Rogers, R., Demidova, E.: {Towards Better
  Understanding Researcher Strategies in Cross-lingual Event Analytics}. In:
  Proceedings of the International Conference on Theory and Practice of Digital
  Libraries. pp. 139--151. Springer (2018)

\bibitem{gottschalk2018eventkgtl}
Gottschalk, S., Demidova, E.: {EventKG+TL: Creating Cross-lingual Timelines
  from an Event-centric Knowledge Graph}. In: Proceedings of the Extended
  Semantic Web Conference. pp. 164--169. Springer (2018)

\bibitem{gottschalk2019eventkg}
Gottschalk, S., Demidova, E.: {EventKG--the Hub of Event Knowledge on the
  Web--and Biographical Timeline Generation}. Semantic Web  \textbf{10}(6),
  1039--1070 (2019)

\bibitem{hecht2010tower}
Hecht, B., Gergle, D.: {The Tower of Babel meets Web 2.0: User-generated
  Content and its Applications in a Multilingual Context}. In: Proceedings of
  the SIGCHI conference on human factors in computing systems. pp. 291--300
  (2010)

\bibitem{kaffee2019ranking}
Kaffee, L.A., Endris, K.M., Simperl, E., Vidal, M.E.: {Ranking Knowledge Graphs
  By Capturing Knowledge about Languages and Labels}. In: Proceedings of the
  10th International Conference on Knowledge Capture. pp. 21--28 (2019)

\bibitem{kaltenbrunner2012there}
Kaltenbrunner, A., Laniado, D.: {There is no Deadline: Time Evolution of
  Wikipedia Discussions}. In: Proceedings of the Eighth Annual International
  Symposium on Wikis and Open Collaboration. pp. 1--10 (2012)

\bibitem{marie2014exploratory}
Marie, N., Gandon, F., Giboin, A., Palagi, {\'E}.: {Exploratory Search on
  Topics through Different Perspectives with DBpedia}. In: Proceedings of the
  10th International Conference on Semantic Systems. pp. 45--52 (2014)

\bibitem{miz2020trending}
Miz, V., Hanna, J., Aspert, N., Ricaud, B., Vandergheynst, P.: {What is
  Trending on Wikipedia? Capturing Trends and Language Biases Across Wikipedia
  Editions}. In: Companion Proceedings of the Web Conference 2020. ACM (2020)

\bibitem{mocanu2013twitter}
Mocanu, D., Baronchelli, A., Perra, N., Gon{\c{c}}alves, B., Zhang, Q.,
  Vespignani, A.: {The Twitter of Babel: Mapping World Languages through
  Microblogging Platforms}. PloS one  \textbf{8}(4) (2013)

\bibitem{nguyen2018trio}
Nguyen, T., Tran, T., Nejdl, W.: {A Trio Neural Model for Dynamic Entity
  Relatedness Ranking}. In: Proceedings of the 22nd Conference on Computational
  Natural Language Learning. pp. 31--41 (2018)

\bibitem{oeberst2016individual}
Oeberst, A., Cress, U., Back, M., Nestler, S.: {Individual Versus Collaborative
  Information Processing: The Case of Biases in Wikipedia}. In: Mass
  collaboration and education, pp. 165--185. Springer (2016)

\bibitem{rupnik2016news}
Rupnik, J., Muhic, A., Leban, G., Skraba, P., Fortuna, B., Grobelnik, M.: {News
  across Languages-cross-lingual Document Similarity and Event Tracking}.
  Journal of Artificial Intelligence Research  \textbf{55},  283--316 (2016)

\bibitem{tran2017beyond}
Tran, N.K., Tran, T., Nieder{\'e}e, C.: {Beyond time: Dynamic Context-aware
  Entity Recommendation}. In: European Semantic Web Conference. pp. 353--368.
  Springer (2017)

\end{thebibliography}

\end{document}